\documentclass[useAMS,usenatbib,usegraphicx]{mn2e}

\title[The density of massive early-types  to $z\simeq1.7$]
{The density of very massive evolved galaxies 
to $z\simeq1.7$\thanks{Based on observations made at the Italian Telescopio 
Nazionale (TNG, www.tng.iac.es) operated on the island of La Palma by the 
Centro Galileo Galilei of the INAF (Istituto Nazionale di Astrofisica).}}
\author[P. Saracco et al.]{P. Saracco$^{1}$\thanks{E-mail:
saracco@brera.mi.astro.it}, M. Longhetti$^{1}$, P. Severgnini$^{1}$,
R. Della Ceca$^{1}$, V. Braito$^{1}$,
\newauthor F. Mannucci$^{2}$, R. Bender$^{3,4}$, N. Drory$^{5}$,
G. Feulner$^{3}$, U. Hopp$^{3}$, 
C. Maraston$^{4}$ \\
$^{1}$INAF - Osservatorio Astronomico di Brera, Via Brera 28, 20121 Milano\\
$^{2}$IRA-CNR, Firenze, Italy\\
$^{3}$Universit\"ats-Sternwarte M\"unchen, Scheiner Str. 1, 81679 
      M\"unchen, Germany\\
$^{4}$Max-Plank-Institut fuer extraterrestrische Physik, Giessenbachstrasse,
85748 Garching, Germany\\
$^{5}$University of Texas at Austin, Austin, Texas 78712
}
\begin{document}

\date{Accepted ***. Received 2004; in original form 2004}
\pagerange{\pageref{firstpage}--\pageref{lastpage}} \pubyear{2004}
\maketitle
\label{firstpage}
\begin{abstract}
We spectroscopically identified 7 {massive evolved galaxies }
with magnitudes 
17.8$<$K$<18.4$ at $1.3<z<1.7$ over an area of $\sim160$ arcmin$^2$
of the MUNICS survey. 
Their rest-frame K-band absolute magnitudes are -26.8$<$M$_K<-26.1$ 
(5L$^*<$L$_K<$10L$^*$) and the resulting stellar masses are in
the range $3\div 6.5\times 10^{11}$ M$_\odot$.
{ The analysis we performed unambiguously shows the early-type  
nature of their spectra.} 
The 7 massive evolved galaxies account for a comoving density of  
$(5.5\pm2)\times10^{-5}$ Mpc$^{-3}$ at $\langle z\rangle\simeq1.5$,
a factor 1.5 lower than the density  ($(8.4\pm1)\times10^{-5}$ Mpc$^{-3}$)
of early-types with comparable masses at $z=0$.
{ The incompleteness ($\sim30$\%)  of our spectroscopic observations
accounts for this  discrepancy. 
Thus, our data  do not support  a   decrease of the 
comoving density of early-type galaxies  with masses 
comparable to the most massive ones in the local Universe 
up to $z\simeq1.7$.
This suggests that massive evolved galaxies do not play an important role
in the  evolution of the mass density  outlined by recent surveys in this 
redshift range,}
evolution which instead has to be ascribed to the accretion of the stellar 
mass in  late-type galaxies.
Finally, the presence of such massive evolved galaxies at these redshifts 
suggests that the assembly of massive spheroids has taken place at $z>2$ 
supporting a high efficiency in the accretion of the stellar mass 
in massive halos in the early Universe.
\end{abstract}


\section{Introduction}
The epoch of formation of high-mass 
($\mathcal{M}_{star}>10^{11}$ M$_\odot$)
early-type galaxies is one of the open questions relevant to the 
whole picture of galaxy formation and evolution.
The uniform properties shown by the local ellipticals suggested
the simple monolithic collapse scenario of galaxy formation (Eggen et al. 
1962; Arimoto \& Yoshii 1987). 
On the other hand, the  recent picture  outlined by  
hierarchical models (White \& Frank 1991; Kauffmann 1996; Somerville \& 
Primack 1999) depicts  the formation of local ellipticals through 
subsequent mergers: the higher the final stellar mass of the galaxy, the later 
it has been assembled.  
The most massive of them ($10^{11}-10^{12}$ M$_\odot$) 
populating the brightest end (L$\gg$L$^*$) of the luminosity function 
of galaxies are those reaching their final mass most recently in this 
scenario, possibly at $z<1$. 
Therefore, looking for $z>1$  early-types with stellar masses
comparable to those of the most massive  local ones 
is one of the most direct ways to address  the question 
of galaxy formation.
This is what we are doing  through  a near-IR spectroscopic 
survey of early-type galaxy candidates selected to be 
at $z>1$ and to have stellar masses well in excess to $10^{11}$ M$_\odot$
(Saracco et al. 2003a,b).
The candidates consist of a complete sample of 31 bright (K'$<$ 18.5) 
Extremely Red Objects (EROs) with colours R-K'$\ge$ 5.3 selected over two 
fields ($\sim$320 arcmin$^2$) of the Munich Near-IR Cluster  Survey  
(MUNICS; Drory et al. 2001).
Here, we report the spectroscopic confirmation for a sample of 7 high-mass 
($\mathcal{M}_{star}>3\times10^{11}$ M$_\odot$) 
field early-type galaxies identified at $1.3 <z< 1.7$. 
Throughout this paper we assume H$_0$=70 km s$^{-1}$
Mpc$^{-1}$, $\Omega_0=0.3$ and $\Lambda_0=0.7$.

\section[] {Sample selection, Observations and spectroscopic Classification}
\subsection[]{Sample selection}
The main aim of our  spectroscopic survey is to identify early-type 
galaxies  at $z>1.2$
having stellar masses comparable to  the most massive early-types in the 
local Universe ($10^{11}-10^{12}$ M$_\odot$).
The red optical-to-near-IR colour (R-K$\ge5.3$) favors the selection of 
$z>1$ passively evolved galaxies while the magnitude  K'$<$ 18.5  
 assures the selection of systems with stellar masses well 
in excess of $\mathcal{M}_{star}>10^{11}$ M$_\odot$.  
Indeed, pushing back in time  a local massive  elliptical (L=2.5L$^*$, 
$2-3\times10^{11}$ M$_\odot$)
it would be observed with a  magnitude brighter than K=18.5  up to 
$z\simeq1.5$. 
{ This is shown in Fig. 1 where the expected K-band apparent magnitude
of  local bright  ellipticals is plotted as a function of redshift
(we considered M$^*_K$=-24.3, Kochanek et al. 2001) in case of  pure passive 
evolution.}
We modeled the elliptical by assuming a Simple Stellar Population (SSP) 
model (Maraston 1998, 2004) 2 Gyr old (Z=Z$\odot$) at $z\sim1.5$.
It is worth of noting that even little star formation would produce
a brightening of the expected apparent magnitude.
Thus, K'$<$18.5 is a stringent  criterion which, together with the colour 
criterion, assures that the resulting early-types at $1.2<z<2$ are 
as massive as the most massive ones ($10^{11}-10^{12}$ M$_\odot$) in
the present-day Universe.
The resulting sample  of EROs comprises 31 galaxies.
Thus, we measured a  surface density of R-K$\ge$5.3 galaxies  
of 0.10$\pm$0.02 arcmin$^{-2}$ at K$<18.5$,  in agreement with  previous estimates
over larger areas (e.g. Daddi et al. 2000; Martini 2001).

\begin{figure}
\centering
\includegraphics[width=8cm,height=7cm]{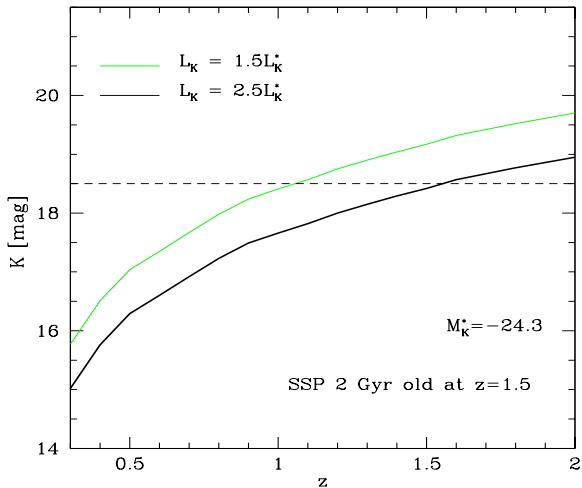}
\caption{ Expected K-band apparent magnitude for an L$_K$=1.5L$^*$ (thin line) 
elliptical and for a L$_K$=2.5$\times$L$^*$ (thick line) 
elliptical as a function of redshift. 
The elliptical has been modeled by assuming a SSP model
based on the evolutionary population synthesis of Maraston (1998, 2004).
The elliptical is 2 Gyr old at $z\simeq1.5$ and evolves passively down 
to $z=0$.  The dashed line marks the selection criterion K$\le$18.5.    
}
\end{figure}

\subsection{Observations}
The spectroscopic observations, with typical exposure of about 4 hours 
for each source, were carried out in  October 2002 and November 2003
at the Italian 3.6 m Telescopio Nazionale Galileo (TNG).
The prism disperser Amici mounted at the near-IR camera NICS of the TNG
was adopted to carry out the observations.
This prism provides the spectrum from 0.85 $\mu$m to 2.4 $\mu$m in
a single shot at a nearly constant  resolution of 
$\lambda/\Delta\lambda\simeq35$ (1.5'' slit width).
This resolution is best suited to describe the spectral shape of 
sources over a wavelength range of $\sim$15000 \AA~ and to detect  
strong continuum features such as the 4000 \AA~ break in
old stellar systems at $z>1.2$.
On the other hand,  the extremely low resolution makes unfeasible the 
detection of emission/absorption lines for sources as faint as our galaxies.
\begin{figure*}
 \centering
 \includegraphics[width=8cm]{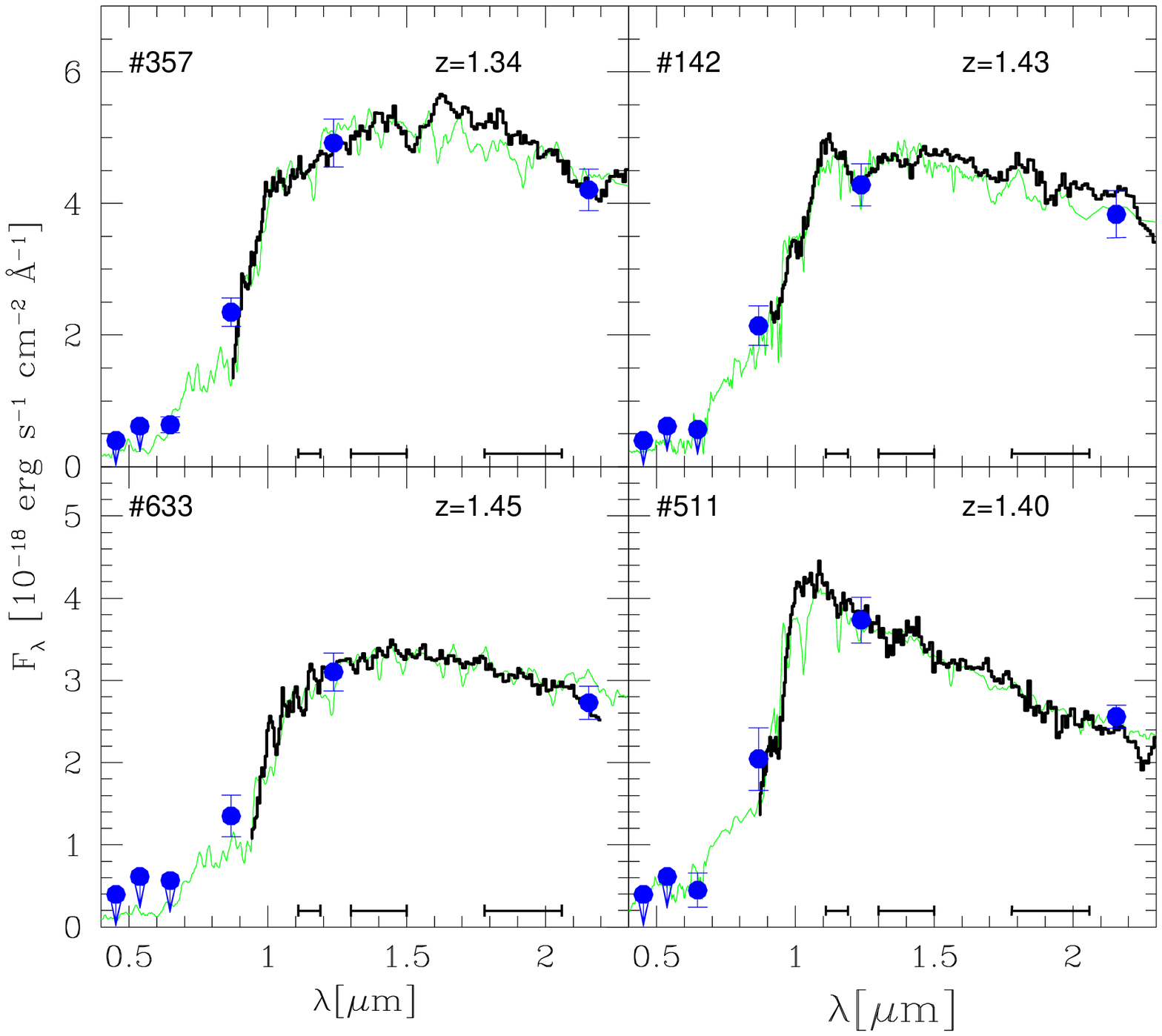}
 \includegraphics[width=8cm]{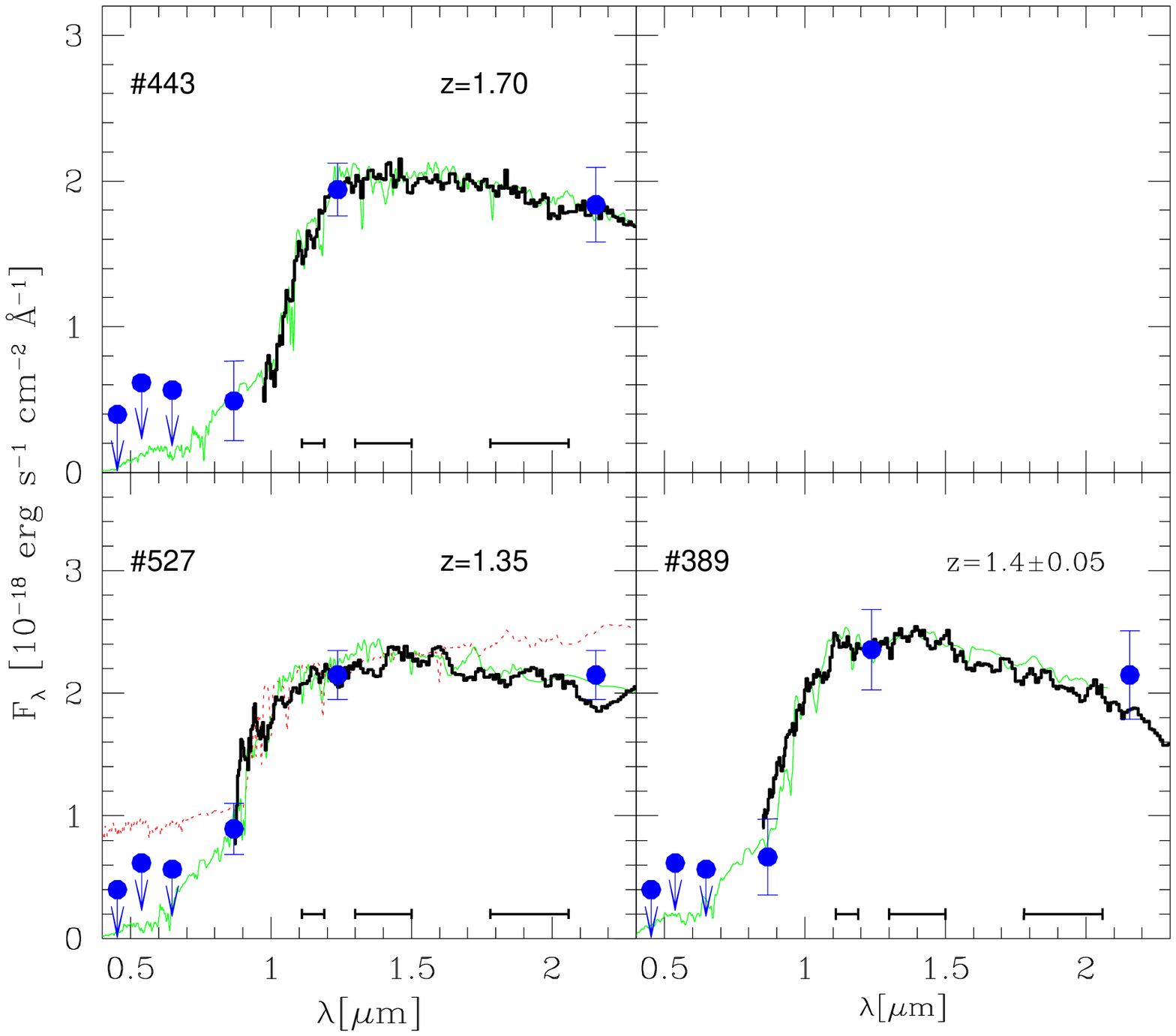}
\caption{Near-IR spectra (black histogram) of 4 out of the 7 early-type 
galaxies. 
The mean observed spectrum of local ellipticals (thin grey line) 
of Mannucci et al. (2001) and of Coleman et al. (1980) are superimposed
on the observed spectrum of S2F1\_357 and of S2F1\_633 and on the
observed spectrum of S2F1\_443 respectively. 
A SSP  1 Gyr old at solar metallicity  is superimposed on 
the observed spectrum of S2F1\_511.
{ For the other galaxies, we superimposed on the Amici spectra
the best-fitting template (see Tab.1).
In the case of S2F1\_527, the best-fitting starburst template is also shown 
(dotted line).}
The horizontal errorbars  represent the atmospheric windows with an 
opacity larger than 80\%. 
The filled symbols are the photometric data in the B, V, R, I, J and K' bands 
from  the MUNICS catalog (Drory et al. 2001). 
}
\end{figure*}
Until now, we carried out spectroscopic observations for $\sim60\%$
of the whole sample identifying 10 early-types.
The analysis of the spectral properties of the whole sample of early-types 
is presented in a forthcoming paper (Longhetti et al. 2004).
Seven of the 10 early-types fall on one of the two selected fields, 
the S2F1 field ($\sim160$ arcmin$^2$), where we collected 13 spectra out 
of the 19 EROs satisfying the selection criteria.
{ In Tab. 1 we report the broad-band photometry of the 7 early-types 
in this field.}

\subsection{Spectral Classification}
{ In Fig. 2  the (smoothed) near-IR spectra (black thick histogram) of 
the 7  galaxies are shown.
The spectra   drops very rapidly at $\sim0.9-1$ $\mu$m concurrent with the 
4000\AA\ break placing the galaxies at $z>1.3$ and suggesting an early-type 
spectral nature.
We searched for the best-fitting template by comparing
the observed SED of each galaxy, constituted by the broad band photometry 
(B, V, R, I, J and K) and by the observed near-IR continuum (from 0.9 $\mu$m
to 2.3 $\mu$m), with a set of spectrophotometric templates  
(Z=Z$\odot$ and Salpeter IMF) based on the Bruzual \& Charlot (2003) 
models and on the library of Simple Stellar Population (SSP) of Maraston 
(1998, 2004).
The best-fitting templates have been obtained through a $\chi^2$ minimization 
applied over the whole wavelength range.
Besides the SSPs, models with declining star formation with time scales 
$\tau$ in the range 0.1-4 Gyr has been considered.
All the observed SEDs are best-fitted by SFHs with very short time
scales ($\tau\le0.3$) Gyr and extinctions E(B-V)$\le0.15$ (with the Calzetti 
et al. (2000) extinction law), providing ages in the range 1.5-4 Gyr.
A comprehensive study of the properties of the stellar populations
in our massive early-types and of their dependence on different IMFs 
and metallicity is presented in Longhetti et al. (2004). 
In Table 1,  the spectroscopic redshift and the best-fitting SFHs  are 
summarized.
It is worth of noting that a set of SSPs with ages in the range 1-3.5 Gyr
provides  a  good fit as well to all the observed SEDs.
To rule out the possibility that some (or all) of the observed SEDs
can be fitted by a young dusty starburst, we forced the fitting procedure 
to search for an acceptable fit among a set of templates made up by
the 6 starburst templates (SB1-SB6) of Kinney et al. (1996) and a starburst
model  described by a constant star formation rate (cst).
Extinction has been allowed to vary in the range 0$<$E(B-V)$<2$ in the 
fitting procedure. 
We did not obtain an acceptable fit 
for any of the 7 galaxies.
In particular, we fail in simultaneously fitting  the red part 
of the spectrum  at $\lambda>1.4~~\mu$m and the blue part at 
$\lambda<0.9~~\mu$m.
This is shown  in Fig. 2, where we plot the best-fitting 
starburst template (dotted line) to the most favorable case
represented by S2F1\_527. 
Thus, dusty starbursts do not reproduce the sharp and deep drop seen in the 
continuum  at 0.85$<\lambda_{obs}<1.0$ $\mu$m.
This confirms the early-type spectral nature  of the 7 galaxies.}
For comparison, in Fig. 2, the mean observed spectrum of local ellipticals
of Mannucci et al. (2001) (thin grey line) is superimposed on the Amici 
spectrum of  S2F1\_357 and of S2F1\_663, while  that of Coleman et al. (1980)
is superimposed on the spectrum of  S2F1\_443.
For the other galaxies, the best-fitting template is superimposed
on the observed spectrum. 
It is worth of noting that, the highest redshift evolved galaxy of
our sample (S2F1\_443) is also one of the X-ray emitting 
EROs we detect on the S2F1 field (Severgnini et al. 2004). 

\section{Luminosities and stellar masses}
The bright K'-band magnitudes (17.8$<$K'$<$ 18.4) of our ellipticals 
and their redshift imply luminosities L$\gg$L$^*$.
Since local L$^*$ galaxies have stellar masses of the order 
of $10^{11}$ M$_\odot$, we expect all the 7 early-types more 
massive than $10^{11}$ M$_\odot$ leaving aside any model assumption.
In order to derive the rest-frame K-band luminosity, the proper 
k-correction of each galaxy has been computed by means of the
relevant best-fitting template.
Each best-fitting template has been multiplied with the transmission
curve of the K filter to derive the k-corrections.
In Tab. 1 we report the K-band absolute magnitudes thus obtained.
As expected, the 7 early-types have rest-frame near-IR luminosities 
5L$^*<$L$_K\le$10L$^*$.
{ The  stellar mass $\mathcal{M}_{star}$ of each galaxy has been 
derived by the K-band luminosity through the mass-to-K-band 
light ratio relevant to the best-fitting template (Z=Z$_\odot$ and Salpeter
IMF) and by the  scale factor applied to the flux of the redshifted template 
to fit the observed fluxes. 
This latter estimate does not consider a certain band but takes into account 
the whole observed SED.
The two estimates provides similar values  (within few percent) since
most of our data  are in the near-IR and the uncertainty on them
are lower than the uncertainty affecting the optical data.
Consequently, the fitting procedure weights mostly the near-IR data. 
An analysis of the dependence of the stellar mass of the 7 evolved galaxies 
on the IMF and on the metallicity is discussed in Longhetti et al. (2004).}   
In Tab.1 we report for each galaxy the rest-frame K-band luminosity, 
the $\mathcal{M}/L_K$ and the relevant stellar mass.
As expected, all the galaxies have stellar masses in the range 
$3\div 6.5\times 10^{11}$ M$_\odot$.
\begin{table*}
\caption{Properties of the 7 early-type galaxies. 
Magnitudes are in the Vega system. { The SFHs, the stellar masses and the 
mass-to-K-band light ratio refer to Z=Z$_\odot$ models with Salpeter IMF.}
}
\centerline{
\begin{tabular}{ccrcccccc}
\hline
\hline
  ID      &   K          & R-K &$z_{spec}$ &M$_K$ & SFH & $\mathcal{M}_{star}$  & M/L$_K$& (L$_{z=0}$/L$^*$)\\
          &              &     &           &   &[Gyr]   & [10$^{11}$ M$_\odot$]& & \\
  \hline
S2F1\_357 & 17.84$\pm$0.08& 6.0   &1.34$\pm$0.05 &-26.6$\pm$0.12 & SSP        & 5.0  & 0.5& 3.0\\
S2F1\_527 & 18.30$\pm$0.15& $>5.7$&1.35$\pm$0.05 &-26.3$\pm$0.20 & $\tau=0.1$ & 3.0  & 0.4& 2.3\\
S2F1\_389 & 18.23$\pm$0.12& 5.5   &1.40$\pm$0.05 &-26.5$\pm$0.15 & $\tau=0.3$ & 3.5  & 0.4& 2.7\\
S2F1\_511 & 18.14$\pm$0.15& 6.1   &1.40$\pm$0.05 &-26.2$\pm$0.20 & $\tau=0.1$ & 3.0  & 0.4& 2.1\\
S2F1\_142 & 17.84$\pm$0.07& 6.0   &1.43$\pm$0.05 &-26.6$\pm$0.12 & $\tau=0.3$ & 6.5  & 0.6& 3.0\\
S2F1\_633 & 18.20$\pm$0.12& $>5.7$&1.45$\pm$0.05 &-26.1$\pm$0.15 & $\tau=0.1$ & 4.0  & 0.6& 1.9\\
S2F1\_443 & 18.40$\pm$0.15& $>5.6$&1.70$\pm$0.05 &-26.8$\pm$0.20 & $\tau=0.1$ & 5.0  & 0.4& 3.6\\
\hline
\hline
\end{tabular}
}
\end{table*}

\section{The density of very massive early-type galaxies}
We estimated the co-moving spatial density of the 7 massive early-type 
galaxies and its statistical uncertainty  as:
\begin{equation}
\rho=\sum_i{1\over V_{max}^i},~~~\sigma(\rho)=\Big[\sum_i\big({1\over V_{max}^i}\big)^2\Big]^{1/2}
\end{equation}
where 
\begin{equation}
V_{max}={\omega\over{4\pi}}\int_{z_1}^{z_{max}}{dV\over{dz}}dz
\end{equation}
is the comoving volume.
The solid angle  $\omega$ subtended by the S2F1 field is
$\sim1.3\cdot 10^{-5}$ strd
and $z_{max}$ is the maximum redshift at which each galaxy
would be still included in the sample.
In the derivation of the $z_{max}$ of each galaxy we computed the 
k-correction by using the relevant best fitting template.
The lower bound in the integration is set to $z_1=1.2$.
It is imposed by the adopted spectroscopic wavelength range which does not 
allow us to detect the Balmer break of galaxies at $z<1.2$.
We find that, at the average redshift $\langle z_{max}\rangle\simeq1.55$, the
7 massive early-types  account for a co-moving density 
$\rho=(5.5\pm2)\times 10^{-5}$ Mpc$^{-3}$ over a volume of about 
$1.5\times 10^{5}$ Mpc$^{3}$.
Given the stellar masses  of the 7 early-types the resulting mass density  
is $\rho_{star}=(2.3\pm0.9)\times 10^7$ M$_\odot$ Mpc$^{-3}$.
Such  densities are likely to be  lower limits  because of the 
incompleteness of our spectroscopic observations 
(13 spectra collected out of the 19 EROs in the field).
{ In order to compare these densities with the local values we integrated 
the K-band LF of local early-type galaxies derived by Kochanek et al. (2001)
described by M$_K^*=-24.3\pm0.06$ and $\Phi^*=1.5\pm0.2\times10^{-3}$ 
Mpc$^{-3}$.} 
The lower bound to the integration has been derived  
in two independent ways.
The first one  lies in  deriving the lowest luminosity that our galaxies
would have at z=0.
This has been obtained by assuming a pure passive evolution from $z_{spec}$
to $z=0$.
{ The luminosities thus obtained are reported in Tab. 1
and place the lower limit at 1.9L$^*$ as imposed by  S2F1\_633.
In the second way, we derived the K-band luminosity from 
the lowest stellar mass of our galaxies ($3\times10^{11}$ M$_\odot$)
by means of the maximum mass-to-light ratio that our galaxies
could have at $z=0$.
Since their stellar populations have to be formed at $z>2$, 
they would be $\sim11$ Gyr old at $z=0$.
Thus, the maximum mass-to-light ratio is $\mathcal{M}$/L$_K\simeq$1.6,
given by a SSP 11 Gyr old with  Salpeter IMF.
Also in this case we obtained a luminosity of about 1.9L$^*$.
It is worth of noting that  other IMFs would produce lower values of 
$\mathcal{M}$/L$_K$ (e.g. $\mathcal{M}$/L$_K$=1.3 with Kroupa IMF and 
$\mathcal{M}$/L$_K$=1.1 with Miller-Scalo) and, correspondingly, higher 
luminosities (2.3L$^*$ and 2.7L$^*$ respectively).
By integrating the local LF of early-type galaxies at luminosities
brighter than 1.9L$^*$ we obtained a density of $(8.4\pm1)\times10^{-5}$ 
Mpc$^{-3}$.}
We summarized our results in Fig. 3 where  we report also the comoving
density we derived for the 3 early-types with L$_{z=0}>3$L$^*$. 
{  The density we estimated at $z\simeq1.5$ represents 65\%  
of the density of their counterparts at $z=0$.
Even if this difference is not statistically significant,
this suggests that  three more massive early-types should be expected  
over  the 160 arcmin$^2$ in case of no density evolution.}
On the other hand, our estimate is affected by a spectroscopic redshift 
incompleteness of about 30\%. 
When this incompleteness is taken into account the
observed difference tends to vanish.
Thus,  we  conclude that the number density of early-types with 
stellar masses comparable to the most massive early-types populating the 
local Universe does not show evidences of  decrease up 
to $z\simeq1.7$.

{ In Fig. 3 we also report the density relevant 
to the four  old spheroidal galaxies (empty circle) spectroscopically 
identified at $z>1.6$  by Cimatti et al. (2004), 
the  density of  massive ($10^{11}$ M$_\odot$) evolved galaxy 
candidates at $z>2$ (open triangle) found by  Saracco et al. (2004) 
on the HDF-S and the density found by Caputi et al. (2004)  
on the GOODS/CDF-S area (crosses), both on the basis of photometric analysis.
Finally,  the upper limit to the number density of galaxies with masses 
larger than $2\times10^{11}$ M$_\odot$ (open squares) by Drory et al. (2004)
is shown.}
\begin{figure}
\centering
\includegraphics[width=8cm]{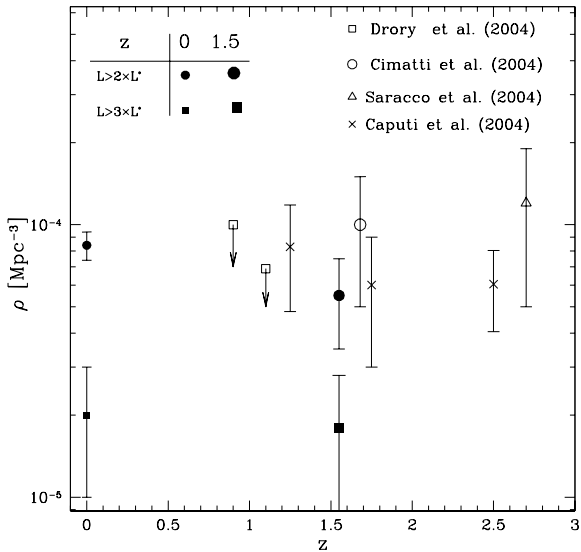}
\caption{Number density of galaxies as a function of redshift.
The large filled symbols represent the densities we derived at 
$z\simeq1.5$ for the 7 early-types with L$_{z=0}>2$L$^*$
(circle) and for the 3 with L$_{z=0}>3$L$^*$ (square).
The small filled symbols at $z=0$ represent the number density
of E/S0 galaxies brighter than 2L$^*$ (circle) and 3L$^*$
(square) respectively obtained by integrating the 
{ local LF
of galaxies of Kochanek et al. (2001).
The open circle is the density of spheroidal galaxies by
Cimatti et al. (2004), the open triangle and the crosses are
the density of $\mathcal{M}>10^{11}$ M$_\odot$ evolved galaxy candidates
by Saracco et al. (2004) and by Caputi et al. (2004) 
respectively and, finally, the open
squares are the upper limit to the number density of galaxies with
$\mathcal{M}>2\times 10^{11}$ M$_\odot$ by Drory et al. (2004).}
}
\end{figure}

\section{Conclusions}
We spectroscopically identified  7 bright ($17.8<$K$<18.4$)
massive evolved galaxies at $0<z<1.7$ over an area of about 160 arcmin$^2$
of the MUNICS survey.
These galaxies turned out to have rest-frame K-band luminosities
5.5L$<$L$_K\le$11L$^*$ and stellar masses in the range 
$3-6.5\times 10^{11}$ M$_\odot$.
At the mean redshift of $z\simeq1.5$ these 7 early-types
sample a volume of about $1.5\times 10^{5}$ Mpc$^{3}$ and
account for a comoving number  density 
$\rho=(5.5\pm2)\times 10^{-5}$ Mpc$^{-3}$ 
and a stellar mass density 
$\rho_{star}=(2.3\pm0.9)\times 10^7$ M$_\odot$ Mpc$^{-3}$.
{ These densities represent 65\% of the values at $z=0$ for 
early-types with comparable mass.
The incompleteness of our spectroscopic observations (30\%)
accounts for this deficiency.}
Thus, our results  show that the number density of the most massive 
early-types in the present-day Universe keeps essentially constant 
down to $z\simeq1.7$.
This suggests that, massive early-types do not take part in  the evolution 
of the stellar  mass density  in the redshift range $0<z<1.7$ and
that the decrease of the stellar mass density detected in this
redshift range by recent surveys  (e.g. Rudnick et al. 2003;
Drory et al. 2004; Fontana et al. 2004) has to be ascribed to the 
accretion of the stellar mass in massive late-type galaxies.  
This qualitatively agrees with the concurrent increase of the 
cosmic star formation rate due to late-type galaxies seen in the same 
redshift range (e.g. Madau et al. 1998).
The high stellar masses and the number density of the 7 massive
evolved galaxies imply that they were fully assembled 
at the observed redshift  pushing their 
formation at  $z>2$, as also suggested by the recent discovery of an evolved 
spheroidal galaxy at $z\simeq1.9$ (Cimatti et al. 2004).
This is in agreement with the results based on the spectral analysis of 
the stellar populations in our sample of massive early-types at $z\simeq1.5$ 
which suggest a formation redshift  $z_f>2$ (Longhetti et al. 2004).
It is worth of noting that  similar results are also derived by  the 
analysis of the absorption lines indices of  local samples 
(Thomas et al. 2002, 2004) which suggest short timescales ($\sim0.4$ Gyr)
and high-z of formation for massive early-types. 
{ Thus, at variance with the expectations of hierarchical models, 
the most massive early-types in the local Universe do not seem to be
the last galaxies to complete their assembly.
The high formation redshift  suggests an high efficiency
in the accretion of the stellar mass of early-types in the early Universe
and an high star formation preferentially in massive haloes.}

\section*{Acknowledgments}
We thank the staff of the TNG for the very good support during the 
observations.
PS acknowledges financial support by the 
{\em Istituto Nazionale di Astrofisica} (INAF). 
This work has received partial financial support from ASI (I/R/037/01,
I/R/062/02) and from the Italian Ministry of University and the Scientific and
Technological Research (MIUR) through grant Cofin-03-02-23.
The MUNICS project is supported by the Deutsche
   Forschungsgemeinschaft, \textit{Sonderforschungsbereich 375,
   Astroteilchenphysik}

\label{lastpage}

\end{document}